\documentclass[apj]{emulateapj}

\shorttitle{Clustering of Fermi-2LAC Blazars}
\shortauthors{Allevato et al.}

\begin{document}

\title{Clustering of $\gamma$-ray selected 2LAC Fermi Blazars}

\author{V. Allevato\altaffilmark{1,2}, 
A. Finoguenov\altaffilmark{1,2}, 
N. Cappelluti\altaffilmark{3,2}}

\altaffiltext{1}{Department of Physics, University of Helsinki, Gustaf H\"allstr\"omin katu 2a, FI-00014 Helsinki, Finland}
\altaffiltext{2}{University of Maryland, Baltimore County, 1000 Hilltop Circle, Baltimore, MD 21250, USA}
\altaffiltext{3}{INAF-Osservatorio Astronomico di Bologna, Via Ranzani 1, 40127 Bologna, Italy}

\begin{abstract}

We present the first measurement of the 
projected correlation function 
of 485 $\gamma$-ray selected Blazars, divided in 175 BLLacertae (BL Lacs) 
and 310 Flat Spectrum Radio Quasars (FSRQs) 
detected in the 2-year all-sky survey by \textit{Fermi}-Large Area Telescope.
We find that \textit{Fermi} BL Lacs and FSRQs 
reside in massive dark matter halos (DMHs) 
with logM$_h$=13.35$^{+0.20}_{-0.14}$ and
logM$_h$ = 13.40$^{+0.15}_{-0.19}$ h$^{-1}$ M$_{\odot}$, respectively, at 
low (z $\sim$ 0.4) and high (z $\sim$ 1.2)
redshift.
In terms of clustering properties, these results suggest that
BL Lacs and FSRQs are similar objects residing in the same dense environment 
typical of galaxy groups,
despite their different spectral energy distribution, power and accretion rate.
We find no difference in the typical bias and hosting halo mass between \textit{Fermi} Blazars and 
radio-loud AGNs, supporting the 
unification scheme simply equating radio-loud objects with misaligned Blazar counterparts.
This similarity 
in terms of typical environment they preferentially live in, 
suggests that Blazars 
preferentially occupy the centre of DMHs,
as already pointed out for radio-loud AGNs. 
This implies, in light of several projects looking for the $\gamma$-ray emission from 
DM annihilation in galaxy clusters, a strong contamination from Blazars 
to the expected signal from DM annihilation.

\end{abstract}


\section{Introduction}
\label{sec:intro}

Blazars are radio-loud Active Galactic Nuclei (AGNs) 
with jets pointing at us rather than in the plane of the sky
(Blandford et al. 1978, Urry \& Padovani 1995 and references
therein). 
They are characterized by the luminous, rapidly variable and polarized
non-thermal continuum emission, extending from radio to 
$\gamma$-ray (GeV and TeV) energies.
They emit most of their electromagnetic output in the 
$\gamma$-ray band and they are the most numerous 
class of extragalactic objects detected by the Large Area Telescope (LAT) 
onboard the \textit{Fermi} satellite (Nolan et al. 2012) 
and by ground-based Cherenkov telescopes (see Hinton et al. 2009 for a review).
The overall spectral energy distribution (SED) of low power 
Blazars is characterized by two broad distinctive humps 
peaking in the UV-soft X-ray band (due to synchrotron emission of
relativistic electrons) and in the GeV-TeV band (due to inverse Compton
scattering of soft photons by the same relativistic electrons).
High power Blazars peak at smaller frequencies (sub-mm and $\sim$MeV).

Blazars are usually separated into low power BL Lacertaes (BL Lacs) and 
high power flat-spectrum radio quasars (FSRQs). BL Lacs are typically completely 
dominated by the jet emission, emitting a double humped synchrotron 
self-Compton spectrum. 
The FSRQs are more complex, showing clear signatures 
of a ‘normal’ AGN disc and broad line region (BLR), unlike the BL Lacs which 
generally show no broad lines or disc emission.
Thus the nature of the accretion flow itself is 
different in BL Lacs and FSRQs, with the latter showing a 
standard disc which is absent from the former. This can be linked to the clear 
distinction in Eddington ratio between BL Lacs and FSRQs, with the BL Lacs all 
consistent with \.m = \.M/\.M$_{Edd} < 0.01$ (where $\eta$\.{M}$_{Edd}$c$^2$ 
= L$_{Edd}$ and the efficiency $\eta$ depends on BH spin) while the FSRQs 
have \.m $>$ 0.01 (see e.g. Ghisellini et al. 2010, 2013).
According to unified schemes, Blazars are  
misaligned counterparts of radio galaxies, 
%That is,
%according to unified schemes, Blazars are simply radio
%galaxies with their radio jets pointing at us rather 
%than in the plane of the sky (Blandford et a. 1978, Urry \& Padovani 1995
%and references therein).
%In detail, the parent population of
with FSRQs made by powerful Fanaroff-Riley type 2 
(FRII) radio-galaxies and BL Lacs related to low power FRI radio galaxies.

As shown in Ackermann et al. (2011), the \textit{Fermi} 
$\gamma$-ray Space Telescope provided one of 
the largest sample of Blazars up to z=3.1, allowing for 
the first time the study of the spatial distribution
of $\gamma$-ray selected AGN in the Universe.
AGN clustering measurements are powerful in providing 
information about the physics of galaxy/AGN formation and evolution, 
the typical environment that AGN preferentially live in 
and to put constraints on the mechanisms that trigger the AGN activity. 
Up to now a large amount of research in the field of clustering 
has been performed in optical, radio and X-ray band. The amplitude 
of the quasar correlation function suggests that optically selected 
quasars are hosted by halos of roughly constant mass, a few 
times 10$^{12}$ M$_{\odot}$ h$^{-1}$, out to z$\sim$3-4
(e.g., Croom et al. 2005, Porciani \& Norberg 2006, 
Myers et al. 2007, Ross et al. 2009, da Angela et al. 2008). 
On the other hand, measurements of the spatial 
distribution of X-ray AGN show that they are located 
in galaxy group-sized DMHs with M$_h$ = 10$^{13-13.5}$ 
M$_{\odot}$ h$^{-1}$ at low ($\sim$0.1) and high ($\sim$1) redshift 
(see Cappelluti, Allevato \& Finoguenov 2012 for a review).
The fact that DMH masses of this class of moderate luminosity 
AGN is estimated to be, on average, larger than those of luminous
quasars, has been interpreted as evidence against cold gas accretion 
via major mergers in those systems (e.g. Allevato et al. 2011; 
Mountrichas \& Georgakakis 2012), and/or as support for multiple 
modes of BH accretion (cold versus hot accretion mode, Fanidakis et al. 2013).

\begin{figure}
\plotone{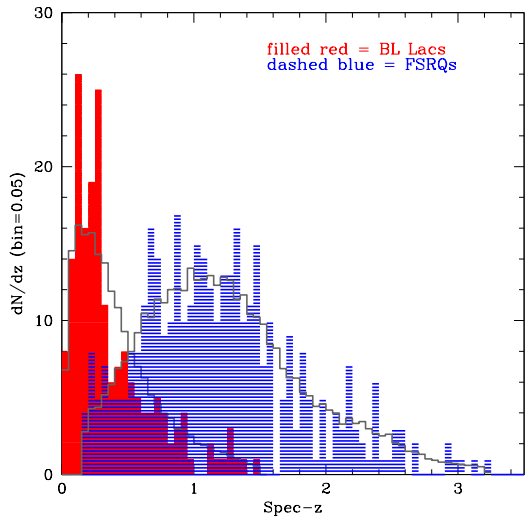}
\caption{\footnotesize Redshift distribution of 175 2LAC BL Lacs (filled red histogram), 310 2LAC FSRQs (dashed blue histogram) and of the corresponding random sources (grey empty histograms), where the random redshifts have been obtained using a Gaussian smoothing with $\sigma$=0.2.}
\label{fig1}
\end{figure}

Compared to X-ray and optically-selected sources, radio-selected 
AGN are more likely found in group/cluster environments 
(Smolcic et al. 2011) and are mainly hosted by massive early-type galaxies
(e.g. Smolcic et al. 2008, 2009).
This result is also confirmed 
by clustering studies showing that 
radio AGNs have the same clustering amplitude
of local elliptical galaxies (e.g., Magliocchetti et al. 2004,
Mandelbaum et al. 2009; Wake et al. 2008, Hickox et al. 2009), 
with a strong dependence on radio luminosity  (e.g., Overzier et al. 2003).
Magliocchetti et al. (2004) found that faint radio-FIRST AGNs
reside in DMHs more massive than 13.4 M$_{\odot}$,
which corresponds (following Ferrare 2002) to a 
threshold BH mass of $\sim$ 10$^9$ M$_{\odot}$.
This minimum mass required to have radio-AGNs
is larger than that obtained for radio-quiet quasars 
in the same redshift range.
Moreover, Hickox at al. (2009) found that radio AGNs reside in DMHs 
with M$_h$ = 3 $\times$ 10$^{13}$ h$^{-1}$ M$_{\odot}$,
while Mandelbaum et al. (2009) estimated a mass
(deduced from g-g lensing) of
1.6 $\times$ 10$^{13}$ h$^{-1}$ M$_{\odot}$ for SDSS radio AGNs.
The radio luminosities of these radio AGNs range from 10$^{23}$ to 
10$^{26}$ W Hz$^{-1}$, i.e. they are mainly FRI systems.

The clustering signal of radio-loud FRI AGNs has been 
also modelled by using the halo occupation 
distribution (HOD). Wake et al. (2008) 
showed that 2SLAQ radio AGNs at 0.4 $<$ z $<$ 0.7,
preferentially occupy central galaxies
respect to satellites. This HOD has been directly confirmed
by Smolcic et al. (2011), who also found that low-power radio 
COSMOS AGNs are preferentially associated with galaxies close to the centre ($<$0.2$R_{200}$).

Clustering of $\gamma$-ray emitting AGN has never been studied
and \textit{Fermi} is the only instrument capable to perform such an analysis. 
Following the argument that Blazars are simply mislagned counterpart 
of radio galaxies, we expect that $\gamma$-ray selected Blazars behave 
similarly to radio-AGN,
i.e. we expect that they are hosted by the most massive DMH halos at each epoch,
with mass greater than few times 10$^{13}$ M$_{\odot}$.

Throughout this work we will assume a $\Lambda$CDM cosmology, 
with $\Omega_M$ = 0.3, $\Omega_{\Lambda}$ = 0.7, 
$\Omega_b$= 0.045, $\sigma_8$ = 0.8. 
All distances are comoving and quoted in units of Mpc h$^{-1}$, 
assuming h = $H_0$/100 km s$^{-1}$. 
The symbol log indicates a base-10 logarithm.

\begin{deluxetable*}{cccccccccc}
\tabletypesize{\scriptsize}
\tablewidth{0pt}
\tablecaption{2LAC Fermi AGN Samples \label{tbl-1}}
\tablehead{
\colhead{(1)} &
\colhead{(2)} &
\colhead{(3)} &
\colhead{(4)} &
\colhead{(5)} &
\colhead{(6)} &
\colhead{(7)} &
\colhead{(8)} &
\colhead{(9)} &
\colhead{(10)} \\
\colhead{Sample} &
\colhead{N} & 
\colhead{$\langle z \rangle$\tablenotemark{a}} &
%\colhead{$\langle L_{\gamma} \rangle$}  &
\colhead{$r_0$} &
\colhead{$\gamma$} &
\colhead{$r_{0,\gamma=1.8}$} &
\colhead{$b_{PL}$} &
\colhead{$b$} &
\colhead{$\frac{\chi_{min}^2}{d.o.f.}$} &
\colhead{logM$_h$} \\
\colhead{} &
\colhead{} &
\colhead{} &
\colhead{Mpc h$^{-1}$} &
\colhead{} &
\colhead{Mpc h$^{-1}$} &
\colhead{$\sigma_{8,AGN}(z)/\sigma_{DM}(z)$} &
\colhead{Halo Model} &
\colhead{} &
\colhead{h$^{-1}$M$_{\odot}$} }
\startdata
%\multicolumn{6}{c}{\textit{Sample 1 - Only spec-zs}}\\
BL Lac & 175 & 0.38 & 6.90$^{+0.34}_{-1.46}$ & 1.64$^{+0.35}_{-0.30}$ & 7.88$\pm0.66$ & 1.52$\pm0.21$ & 1.84$\pm0.25$ & 6.3/8 & 13.35$^{+0.20}_{-0.14}$\\
%(5.56$\pm$0.7)\tablenotemark{*} & 
%& 1.51$\pm$0.35 
%& 4.13$\pm$0.60
FSRQ & 310 & 1.18 & 7.7$^{+3.8}_{-3.1}$ & 1.5$^{+0.3}_{-0.4}$ & 11.2$\pm1.2$ & 3.0$\pm$0.3 & 3.30$\pm0.41$ & 22.3/8 & 13.40$^{+0.15}_{-0.19}$ \\
%\multicolumn{6}{c}{\textit{Sample 2 - Spec \& phot-z}}\\
%All AGN & 348 & 2.79 & $10^{44.92}$ & 3.55$\pm$0.45 & 12.20$^{+0.17}_{-0.25}$\\
%Type 1 & 173 & 2.77  & $10^{45.16}$ & 5.32$\pm$0.63 & 12.80$^{+0.15}_{-0.18}$ \\  
%Type 2 & 175 & 2.81 & $10^{44.70}$ & 2.67$\pm$0.82 & 11.64$^{+0.50}_{-0.71}$\\
%(3.21$\pm$0.72)\tablenotemark{*} & 
\enddata
%\table
%\tablenotetext{a}{Median redshift of the sample.}
%\tablenotetext{b}{Median redshift of the sample.}
%\tablenotetext{*}{$\chi_{min}^2$ is the minimum value assumed by the $\chi^2$ statistic while $d.o.f.$ gives the number of degrees of freedom.}
\tablecomments{Values of $r_0$, $\gamma$ and $r_{0,\gamma=1.8}$ are obtained from a power-law fit of the 2PCF over the range $r_p$=1-80 Mpc h$^{-1}$, using the full error covariance matrix and minimizing the correlated $\chi^2$ values. The bias parameters, $b_{PL}=\sigma_{8,AGN}(z)/\sigma_{DM}(z)$, are based on the power-law best fit parameters $r_{0,\gamma = 1.8}$ and the uncertainties are derived from the standard deviation of $\sigma_{8,AGN}(z)$, where the 1$\sigma$ errors on $\sigma_{8,AGN}(z)$ correspond to $\chi^2=\chi_{min}^2+1$.
The bias factors in col (8) are estimated using the halo model, $w_{mod}(r_p) = b^2w_{DM}(r_p,z)$ where $w_{DM}(r_p,z)$ is the dark matter 2PCF at large scale (2-halo term) evaluated at the mean redshift of the samples.
The 1$\sigma$ errors on the bias correspond to $\chi^2 = \chi^2_{min} + 1$ where the $\chi_{min}^2$ is given in col (9). The correlations between errors have been taken into account through the inverse of the covariance matrix.  To derive logM$_h$ we followed the bias-mass relation b(M$_h$,z) described in van den Bosch (2002) and Sheth et al. (2001), using the bias factors in col (8).} 
\end{deluxetable*}

\begin{figure*}
\plottwo{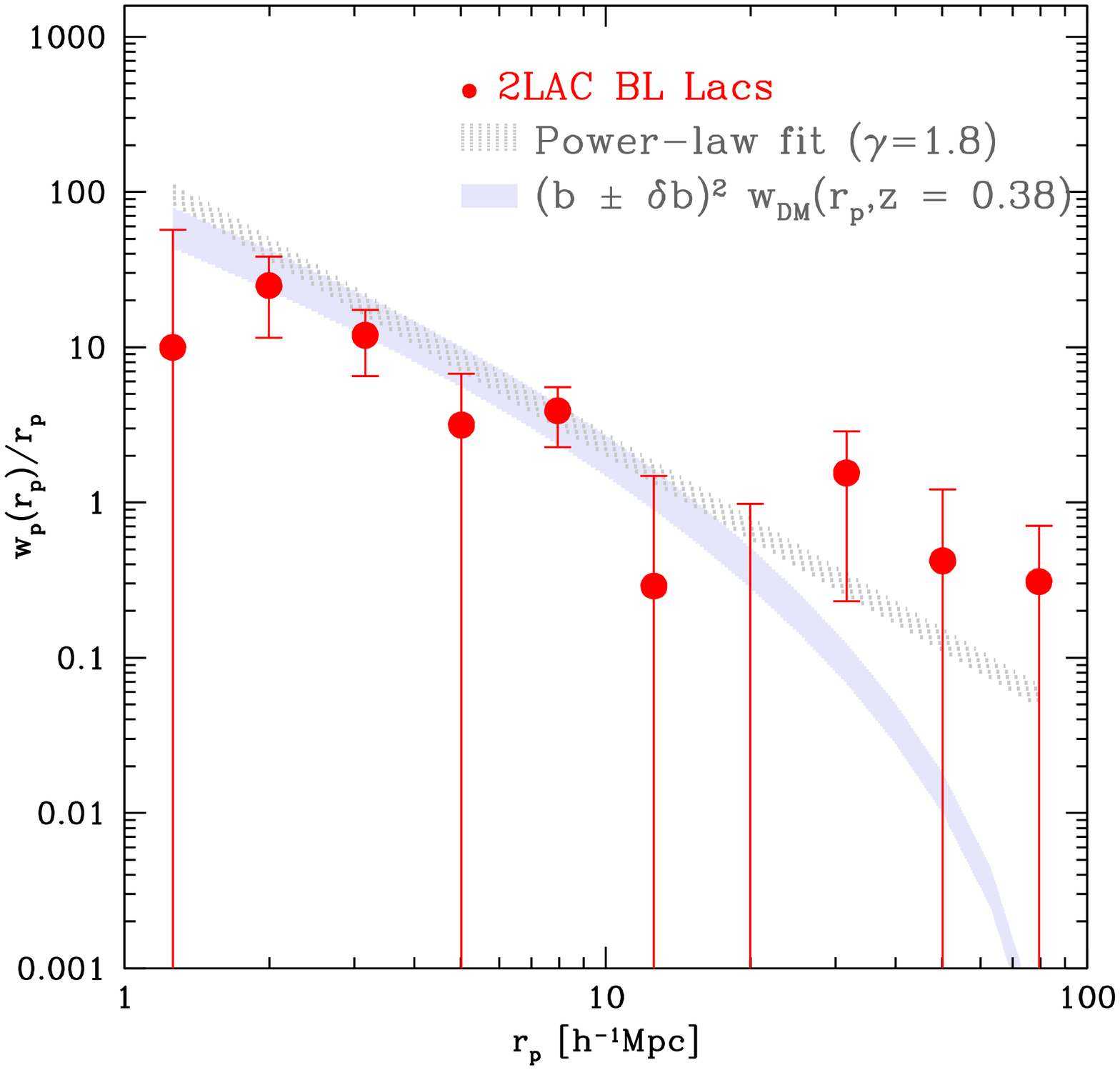}{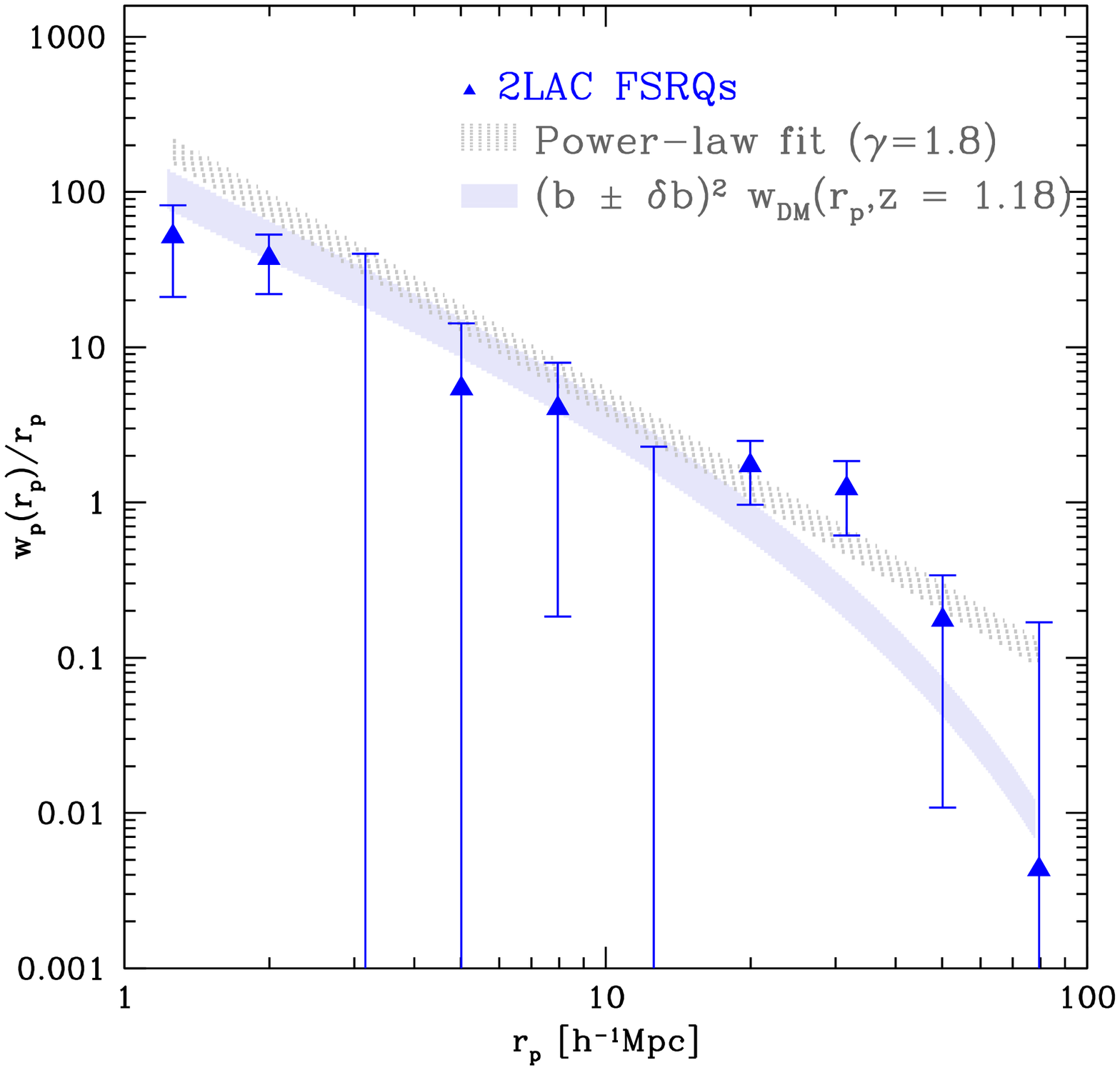}
\caption{\footnotesize Projected 2PCF of 175 2LAC BLLacs (red circles) and 310 FSRQs (blue triangles) compared to $(b \pm \delta b)^2 w_{DM}(r_p)$ (blue shaded region), where $(b \pm \delta b)^2$ is the square of the bias factor $\pm 1\sigma$ (Table 1, col 8) and $w_{DM}(r_p)$ is the dark matter 2-halo term evaluated at the mean redshift of the sample. For comparison, the dotted grey regions mark the best-fit power-laws with $\gamma=1.8$ and $r_0$ given in Table 1, col 6.} 
\label{fig1}
\end{figure*}

\section{AGN Catalog}\label{sec:AGNcat}

The second catalog of AGNs (2LAC) detected by \textit{Fermi}-
Large Area Telescope (LAT) in two years of scientific 
operation has been presented in Ackermann et al. (2011). The entire 
sample includes 1017 $\gamma$-ray sources. 
For each source they found a counterpart comparing the sample with 
known source catalogs and with uniform surveys in the radio and 
X-ray bands.
The associations have been done with statistical approaches 
such as Bayesian, likelihood ratio and logN-logS association methods
(see Ackermann et al. 2011 for more details).
The 1017 LAT sources have been classified in 
FSRQ, BL Lac object, radio galaxy, steep-spectrum radio quasar (SSRQ), Seyfert, 
narrow line Seyfert 1, starburst galaxy. The ingredients of the
classification are optical spectra or other Blazar characteristics 
(radio loudness, broadband emission, variability and polarization)
and the synchrotron-peak frequency of the broadband SED. 
%%through optical spectra with a good evaluation of emission lines.
All the sources without a good optical spectrum or without an optical spectrum at all
are defined as AGU and AGNs.
%(AGNs are not confirmed Blazars nor 
%Blazar candidates, such as AGU).}
All the objects with single counterparts and without analysis flag (886/1017)
define the Clean Sample, which comprises 395 BL Lacs and 310 FSRQs,
157 sources of unknown type, 22 other AGNs and 2 starburst galaxies. 
We focus our analysis on objects classified as Blazars, i.e. BL Lacs and FSRQs, 
with known spectroscopic redshift. 
Unfortunately only 46\% of BLLacs (175/395) have measured redshifts which extend to z=1.5. 
Otherwise the redshift distribution of FSRQs (310) peaks around z=1 and extends to z=3.10.
The redshift distributions of 175 2LAC BL Lacs and 310 2LAC FSRQs 
are compared in Fig. 1. The mean redshift is $\langle z \rangle$=0.38 and 1.18
for BL Lacs and FSRQs, respectively.

\section{2-point correlation function}
\label{sec:2pcf}

The two-point correlation function (2PCF) $\xi(r)$
is defined as the excess probability $dP$ above a 
Poisson distribution of finding an object in a volume element 
$dV$ at a distance $r$ from another randomly chosen object (Peebles 1980). 
With a redshift survey, we cannot directly measure $\xi(r)$ in 
physical space because peculiar motions
of galaxies distort the line-of-sight distances inferred from redshift. 
To separate the effects of redshift distortions, 
%due to peculiar velocities along the line-of-sight, 
the 2PCF is measured in two dimensions, $r_p$ and $\pi$
which are the projected comoving separations between AGN pairs 
in the directions perpendicular and parallel to the line-of-sight, 
respectively. Following Davis \& Peebles (1983)
$r_p$ and $\pi$ are defined as:
\begin{eqnarray}
\pi  & = & \frac{\textbf{s} \cdot \textbf{l}}{\vert \textbf{l} \vert}  \\
r_p  & = & \sqrt{(\textbf{s} \cdot \textbf{s} - \pi^2)}
\end{eqnarray}
where $r_1$ and $r_2$ are the redshift positions of a 
pair of AGN, $s$ is the redshift-space separation 
$(r_1-r_2)$ and $l=(r_1+r_2)/2$ is the mean distance 
to the pair.
We then measure the 2PCF on a two-dimensional grid of 
separations r$_p$ and $\pi$, obtaining the projected correlation 
function w$_p$(r$_p$) defined by Davis \& Peebles (1983) as:
\begin{eqnarray}\label{eq:integral}
w_{AGN}(r_p) = 2 \int_0^{\pi_{max}} \xi(r_p,\pi) d\pi 
\end{eqnarray}
Usually $\xi(r_p,\pi)$ is measured using the estimator defined
in \citet[LS]{Lan93}:
\begin{equation}\label{eq:LZ}
\xi = \frac{1}{RR} [DD-2DR+RR]
\end{equation}
The LS estimator is then defined as the ratio between AGN pairs
in the data sample and pairs of sources in the random catalog, 
as a function of the projected comoving separations between
the objects. This estimator has been used to measure the 2PCF of
X-ray, optically and radio selected AGN (e.g. Zehavi et al. 2005,
Li et al. 2006, Coil et al. 2009, Hickox et al. 2009, 
Gilli et al. 2009, Allevato et al. 2011).

The choice of $\pi_{max}$ is a compromise between having
an optimal signal-to-noise ratio and reducing the excess noise
from high separations. 
Usually, the optimum $\pi_{max}$ value can be determined by
estimating $w_p(r_p)$ for different values of $\pi_{max}$ and finding 
the value at which the 2PCF levels off. Following this
approach, we fixed $\pi_{max}$ = 40 h$^{-1}$Mpc in the following analysis
which ensures the $w_p(r_p)$ convergence.

The measurement of the 2PCF 
required the construction of an AGN random catalog 
with the same selection criteria and observational effects as the data. 
This random sample serve as an unclustered distribution 
to which to compare the data.
We separately created a random catalog
for BLLacs and FSRQs, 
reproducing the space and flux distributions of 2LAC Blazars.
In detail, the random sources are randomly placed in the sky and the fluxes randomly 
drawn from the catalog of real fluxes 
and kept in the random sample if above the values of the sensitivity map 
(published by Abdo et al. 2010) at those random positions. 
We prefer this method with respect to the one that keeps 
the angular coordinates unchanged and then has the disadvantage of removing 
the contribution to the signal due to angular clustering.
The redshifts are randomly extracted from a smoothed redshift 
distribution of the real sources. Specifically, we assume a Gaussian 
smoothing length $\sigma_z = 0.2$. This is
a good compromise between scales 
that are too small, which would suffer from 
local density variations, and those that are too 
large, which would oversmooth the distribution.
However, we verified that our results do not change 
significantly using $\sigma_z$ = 0.2-0.4.
Fig 1 compares the redshift distribution of 175 BL Lacs (in red)
and 310 FSRQs (in blue) with the redshift distribution of the random sources, 
where the random redshifts have been obtained 
using a Gaussian smoothing with $\sigma$=0.2.

Since adjacent bins in w$_p(r_p)$ are correlated, as are their errors,
we constructed the covariance matrix M$_{i,j}$ (Miyaji et al. 2007), which reflects 
the degree to which bin $i$ is correlated with bin $j$.
The covariance matrix is used to obtain reliable fits to w$_p$(r$_p$) 
by minimizing the correlated $\chi^2$ values. 
By using a bootstrap resampling technique (Coil et al. 2009; 
Hickox et al. 2009; Krumpe et al. 2010; Cappelluti et al. 2010), 
we estimated the covariance matrix as: 
\begin{eqnarray} 
M_{i,j} & = & \frac{1}{N_{boot}} \sum_{k=1}^{N_{boot}} \left( w_k(r_{p,i})- \langle w(r_{p,i}) \rangle \right) \\
& \times & \left( w_k(r_{p,j}) - \langle w(r_{p,j}) \rangle \right)
\end{eqnarray}
% \left[ \sum_{k=1}^{N_{boot}} \left( w_k(r_{p,i})- \langle w(r_{p,i}) \rangle \right) 
%\times \left( w_k(r_{p,j}) - \langle w(r_{p,j}) \rangle \right) \right] 
%($\sigma_i$ = $\sqrt{M_{i,i}}$). 
We calculate w$_p$(r$_p$) N$_{boot}$ = 100 times and $w_k(r_{p,i})$ and $w_k(r_{p,j})$ are from the k-th bootstrap. $\langle w(r_{p,i}) \rangle$
and $\langle w(r_{p,j}) \rangle$ are the averages over all the bootstrap 
samples.

In the halo model approach, the 2PCF is modelled 
as the sum of contributions from AGN pairs within individual DMHs 
(1-halo term, $r_p \lesssim 1$ Mpc h$^{-1}$) and in different 
DMHs (2-halo term, $r_p \gtrsim 1$ Mpc h$^{-1}$). 
The superposition of both the terms describes the 
shape of the 2PCF better than a simple power-law.
In this context, the bias parameter 
$b$ reflects the amplitude of the AGN large-scale clustering 
(2-halo term) relative to the underlying DM distribution, i.e.: 
\begin{equation}\label{eq:b}
w_{mod}(r_p) = b^2 w_{DM}^{2-h}(r_p,z)
\end{equation}
The DM 2-halo term is measured at the mean redshift of the sample, using:
\begin{equation}\label{eq:DM}
w_{DM}^{2-h}(r_p)=r_p \int_{r_p}^{\infty} \frac{\xi^{2-h}_{DM}(r)rdr}{\sqrt{r^2-r_p^2}}
\end{equation}
where
\begin{equation}\label{eq:2-halo}
\xi^{2-h}_{DM}(r)=\frac{1}{2\pi^2}\int P^{2-h}(k)k^2 \left[ \frac{sin(kr)}{kr} \right]  dk
\end{equation}
$P^{2-h}(k)$ is the linear power spectrum (Efstathiou, Bond \& White 1992),
assuming a power spectrum shape parameter $\Gamma = 0.2$ which corresponds to
$h=0.7$ (see Seljak 2000, Hamana et al. 2002).
%For comparison, we also fitted the 2PCF 
%with a power-law model of the form
%\begin{equation}\label{eq:PL}
%w_p(r_p) = \frac{( r_0 / r_p )^{-\gamma} (\Gamma(1/2)\Gamma[(\gamma-1)/2] )}{(\Gamma(\gamma/2) )}
%\end{equation} 
%where $\Gamma$(x) is the Gamma function and $\gamma$ and 
%r$_0$ are the free parameters. 
%The power-law best-fit parameters are related to 
%$\sigma_{8,AGN}(z)$ (Peebles 1980):
%The power-law best-fit parameters are related to 
%$\sigma_{8,AGN}(z)$ (Peebles 1980):
%\begin{equation}
%(\sigma_{8,AGN}^2) = J_2 (\gamma) \left( \frac{r_0}{8 Mpc h^{-1}}\right)^2 
%\end{equation}
%where $J_2(\gamma)=72/[(3-\gamma)(4-\gamma)(6-\gamma)2^{\gamma}]$.
%$\sigma_{8,AGN}(z)$ represents the rms fluctuations within 
%a sphere with a co-moving 
%radius of 8 Mpc,
%which is commonly used to express the clustering strength.
%Following this argument, the bias factor is given by: 
%\begin{equation}
%b_{PL}=\sigma_{8,AGN}(z)/\sigma_{DM}(z)
%\end{equation}
%where $\sigma_{DM}(z)$ is evaluated at 8 Mpc h$^{-1}$ and
%normalized to a value of $\sigma_{DM}(z=0)=0.8$.
	
\section{Results}
\label{sec:results}

%Fig. 1 (Right Panel) shows the 2PCF $w_p(r_p)$ in the range $r_p$ = 1-80 h$^{-1}$ Mpc 
%for 175 BL Lacs and 310 FSRQs compared with
%the DM 2PCF at large scale estimated at z = 0.
The projected 2PCFs of 2LAC BL Lacs and FSRQs are shown in Fig. 2 
in the range $r_p$=1-80 h$^{-1}$ Mpc.
The 1$\sigma$ errors on $w_p(r_p)$ are the square root of the 
diagonal components of the covariance matrix.
We find a significant signal at almost all the sampled scales for 
both BL Lacs and FSRQs, even if with relatively large uncertainty 
due to the small number of AGN pairs at each separation. 
On the contrary, we observe a lack of AGN pairs at smaller
separation ($r_p \lesssim 1$ Mpc h$^{-1}$).

We derive the best-fit bias for 2LAC Blazars
by using a $\chi^2$ minimization technique with 1 free parameter,
where $\chi^2 = \Delta^T M^{-1}_{cov} \Delta$.
In detail, $\Delta$ is a vector composed of $w_{AGN}(r_p)-w_{mod}(r_p)$ (see Eq. \ref{eq:integral} and \ref{eq:b}), $\Delta^T$
its transpose and M$_{cov}$ is the covariance matrix.
The dark matter 2PCF, as defined in Eq. \ref{eq:DM},
is evaluated at the mean redshift of the samples, i.e.
$\langle z \rangle$=0.38 and $\langle z \rangle$=1.18 for
BL Lacs and FSRQs, respectively.
For BL Lacs we find a linear bias of 1.84$\pm0.25$ at $\langle z \rangle$=0.38
while for FSRQs we obtained b=3.30$\pm0.41$ at $\langle z \rangle$=1.18
(see Table 1). The errors correspond to $\Delta \chi^2$ = 1. 

For comparison, we also fit the 2PCF with a power-law model of the form 
$w_p(r_p) = ( r_0 / r_p )^{-\gamma} (\Gamma(1/2)\Gamma[(\gamma-1)/2] ) / (\Gamma(\gamma/2) )$ 
(Coil et al. 2009, Hickox et al. 2009, 
Gilli et al. 2009, Krumpe et al. 2010,2012, Cappelluti 
et al. 2010), 
using a $\chi^2$ minimization technique, with $\gamma$ and $r_0$
as free parameters. As shown in Table 1, we derive $\gamma$=1.64$^{+0.35}_{-0.30}$, 
r$_0$=6.90$^{+0.34}_{-1.46}$ Mpc h$^{-1}$ for BL Lacs and 
$\gamma$=1.5$^{+0.3}_{-0.4}$, r$_0$=7.7$^{+3.8}_{-3.1}$ Mpc h$^{-1}$
for FSRQs. The best-fit values of the power-law slope are 
consistent with several results on the clustering of
X-ray selected AGNs (e.g. Coil et al. 2009, Krumpe et al. 2010, Cappelluti et al. 2010) and
optically selected 2QZ luminous quasars (Porciani et al. 2004).
On the other hand, we argue that the low power-law slope observed for
2LAC Blazars might be due to the lack of clustering signal at small
scale ($r_p \lesssim 1$ Mpc h$^{-1}$).

Finally, fixing $\gamma$ = 1.8, we find that BL Lacs 
at $\langle z \rangle \sim$0.4 have a correlation length 
of $r_0$=7.88$\pm0.66$ Mpc h$^{-1}$
while FSRQs at $\langle z \rangle \sim$1.2, 
have a larger correlation length
of $r_0$=11.2$\pm1.2$ Mpc h$^{-1}$. 
At similar redshift ($z \sim 1.4$), 
%X-ray AGNs show lower r$_0$ values
%for $\gamma \sim$ 1.8 (Gilli et al. 2009, Allevato et al. 2011, 
%Yang et al. 2006, Coil et al. 2007).
Shanks et al. (2011) derived a lower best-fit 
r$_0$ = 5.90 $\pm$ 0.14 Mpc h$^{-1}$ (and $\gamma$=1.8) for 
quasars in SDSS DR5, 2SLAQ and 2QZ surveys. 
%of r$_0$ = 5.90 $\pm$ 0.14 Mpc h$^{-1}$, which 
%differs from the value derived for 2LAC FSRQs 
%at 8$\sigma$ level.

The power-law best-fit parameters are related to 
$\sigma_{8,AGN}(z)$, i.e. the rms 
fluctuations within a sphere with a co-moving 
radius of 8 h$^{-1}$Mpc (Peebles 1980):
\begin{equation}\label{eq:bias}
(\sigma_{8,AGN})^{2} = J_2(\gamma)(\frac{r_0}{8 Mpc/h})^{\gamma}
\end{equation}
where $J_2(\gamma)=72/[(3-\gamma)(4-\gamma)(6-\gamma)2^{\gamma}]$. 
Following this argument, we further derive the 
bias $b_{PL}=\sigma_{8,AGN}(z)/\sigma_{DM}(z)$
where $\sigma_{DM}(z)$ is evaluated at 8 Mpc h$^{-1}$ and
normalized to a value of $\sigma_{DM}(z=0)=0.8$.
As shown in Table 1, fixing the power-law slope
$\gamma$=1.8, we find a bias factor for 2LAC BL Lacs and FSRQs 
quite consistent with the values derived using the halo model approach.
The uncertainties on the bias factors are derived 
from the 1$\sigma$ errors on $\sigma_{8,AGN}(z)$, which correspond to 
$\chi^2=\chi^2_{min}+1$.

The projected 2PCF is always obtained by integrating out to some
finite $\pi_{max}$ rather than to infinity.
Van den Bosch 2013 demonstrates that this finite integration range 
introduces errors on the largest scales probed by
the data ($\sim$20 h$^{-1}$Mpc). This is due to the fact 
that the Kaiser effect (coherent galaxy motion that causes an 
apparent contraction of structure along the line
of sight in redshift space, Kaiser 1987) can not be ignored on scales 
$>\pi_{max}$.
In order to estimate the error introduced in our 2PCF measurements 
by the finite integration range, we use the correction factor derived
in Van den Bosch 2013 (see Equation 48 and Figure 6 therein) for
$\pi_{max}$=40 h$^{-1}$Mpc and by integrating over all halo masses.

We find that, without the correction, we underestimate 
the projected 2PCF by $\sim$35\% at r$_p$=20 h$^{-1}$Mpc
for both 2LAC BL Lacs and FSRQs. However this error only slightly affects
the bias factors.  
%based 
%introduces a small 
%change in the bias factors 
%whether on the power-law best fit parameters (col 7 Table 1) or the halo model 
(col 8 Table 1). 
In fact, by introducing the correction factor 
the bias factors of BL Lacs and FSRQs (based on power-law best fit 
parameters or halo model) increase by $\sim$6\%, i.e. within the 
statistical error of $\sim$14\%.
In agreement with Van den Bosch 2013, the result suggests that 
this effect is important when using projected 
correlation functions to constrain cosmological parameters. However, it only 
introduces a small 
change in the bias of relatively small AGN samples.

Finally, we use the bias factor to estimate the typical DMH mass 
hosting the different Blazar samples (under the assumption that
the bias only depends on the halo mass). To this end, we
followed the bias-mass relation $b(M_h, z)$ defined 
by the ellipsoidal collapse model of 
Sheth et al. (2001) and the analytical approximation of van den Bosch (2002).
Table 1 shows the typical DMH
mass of 2LAC Blazars derived by using the bias as
defined in the halo model (col 8).
We found that BL Lacs and FSRQs reside in massive DM halos with 
logM$_h$=13.35$^{+0.20}_{-0.32}$ and logM$_h$=13.40$^{+0.15}_{-0.25}$ h$^{-1}$M$_{\odot}$,
respectively. These results infer for the first time that $\gamma$-ray 
selected AGN reside in group-sized halos at low ($\sim$0.4) and 
at high ($\sim$1.2) redshift.
%Therefore Fermi-Blazars trace the most massive 
%DMHs at z$\sim$1.2. 

\begin{figure}
\plotone{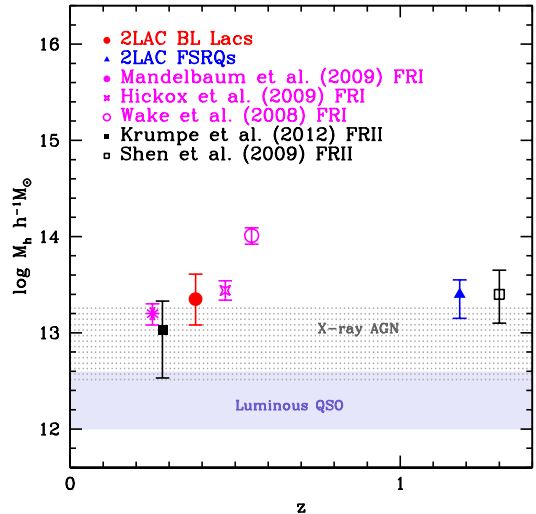}
\caption{\footnotesize Typical DMH mass of 2LAC BL Lacs at $\langle z \rangle$=0.38 (red circle) and 2LAC FSRQs at $\langle z \rangle$=1.18 (blue triangle), estimated following the bias-mass relation b(M$_h$,z) described in van den Bosch (2002) and Sheth et al. (2001). The magenta data points mark the halo mass of radio-loud FRI objects from SDSS (Mandelbaum et al. 2009, deduced from g-g lensing), AGES (Hickox et al. 2009, derived using the bias-mass relation b(M$_h$,z) of Shen et al. 2001) and 2SLAQ LRG survey (Wake et al. 2008, based on the halo occupation distribution). The black data points show the halo mass of radio loud broad line SDSS quasars as estimated in Krumpe et al. (2012) and Shen et al. (2009) derived using the bias-mass relation of Sheth et al. (2001). The quoted typical DMH masses are scaled to the same cosmology ($\Omega_M=0.3$, $\Omega_{\Lambda}=0.7$ and $\sigma_{8}=0.8$). The halo mass evolution with redshift is shown for comparison for optically selected quasars (shaded region, Croom et al. 2005, Porciani et al. 2004, Myers et al. 2006, Shen et al. 2009, Ross et al. 2009, da Angela et al. 2008), and X-ray selected AGN (dotted shaded region, Cappelluti et al. 2010, Allevato et al. 2011, Krumpe et al. 2010, 2012, Hickox et al. 2009, Mountricas et al. 2013, Koutoulidis et al. 2012).} 
\label{fig2}
\end{figure}

\section{Discussion and conclusions}
\label{sec:conc} 

We have used a sample of 485 $\gamma$-ray selected 2LAC Blazars 
(175 BL Lacs and 310 FSRQs) detected in the 2-year all-sky 
survey of the \textit{Fermi} satellite, 
to measure the clustering amplitude and 
to estimate characteristic DM halo masses.
We find that BL Lacs and FSRQs inhabit massive
DMHs with logM$_h$=13.35$^{+0.20}_{-0.14}$ and 
logM$_h$ = 13.40$^{+0.15}_{-0.19}$ h$^{-1}$ M$_{\odot}$,
respectively, at low (z $\sim$ 0.4) and high
(z $\sim$ 1.2) redshift.
%are more biased (4.15$^{+0.55}_{-0.60}$) and reside
%in cluster-sized halos (logM$_h$ = 13.68$^{+0.25}_{-0.20}$
%h$^{-1}$ M$_{\odot}$). 
Usually, BL Lacs and FSRQs have different power, SED and Eddington 
ratio. 
In detail, low power BL Lacs are 
consistent with \.m $< 0.01$, while high power FSRQs 
have accretion rates above this value 
(see e.g. Ghisellini et al. 2010, 2013).
In terms of clustering properties, our results suggest
that BL Lacs and FSRQs are similar objects 
preferentially residing in the same 
dense environment typical of galaxy groups. Additionally, 
their different power and accretion rate, 
that do not translate into different typical bias and halo mass.

It is worth to compare our results inferred for \textit{Fermi}-Blazars
with previous works on radio-loud galaxies.
Radio galaxies are usually classified as 
FRI or FRII sources depending on their radio 
morphology, radio power and optical luminosity
of the hosting galaxies (Fanaroff \& Riley (1974), Ledlow \& Owen 1996). 
Unification models usually associate 
high luminosity FRII with FSRQs 
and low-luminosity FRI with BL Lacs.
%The Fanaroff-Riley classification is somewhat subjective; although most objects can be assigned without ambiguity to one or the other classes, there are a number of intermediate objects which are difficult to classify ([50]). FR Is generally have a 178 MHz luminosity below ~ 2.5 1033 erg s-1 Hz-1 while FR IIs are stronger than this value ([117]); the division in power between the two classes is however not sharp, but when radio sources are plotted as points in the radio luminosity-optical luminosity plane, the Fanaroff-Riley break becomes very sharp, with the break radio power approximately proportional to the optical luminosity squared

Fig. 3 shows the DMH mass of radio-loud SDSS quasars at 0.01 $<$ z $<$ 0.3
(Mandelbaum et al. 2009), 2SLAC quasars at z = 0.55 (Wake et al. 2008)
and radio AGES AGNs at z = 0.57
(Hickox et al. 2009). The radio luminosities of these samples 
range from 10$^{23}$ to 10$^{26}$ W Hz$^{-1}$, 
i.e. typical luminosities of FRI systems. 
Mandelbaum et al. (2009) measured the shear 
signal due to galaxy-galaxy lensing 
and inferred that radio SDSS AGNs reside in DMHs with typical 
mass of 1.6$(\pm0.4)\times$ 10$^{13}$ h$^{-1}$ M$_{\odot}$ at z = 0.25.
This value is quite consistent with the typical mass obtained
for \textit{Fermi}-BL Lacs (M$_h$= 2.7$(\pm1.3)\times$ 10$^{13}$ h$^{-1}$ M$_{\odot}$).
%studied the clustering properties and . In fact, using g-g lensing, they find that radio
%(i.e. radio AGNs reside in more massive DMHs than 
%AGNs without radio emission and galaxies without the presence of AGN activity).
Hickox at al. (2009) found similar typical mass of the 
hosting halos for radio AGES AGNs
(M$_h$ = 3$(\pm0.9)\times$10$^{13}$ h$^{-1}$ M$_{\odot}$).
%Our measurements are also in agreement with the results shown in Magliocchetti et al. (2004).
%They found that radio FIRST galaxies with AGN activity (mainly FRI objects)
%reside in DMHs more massive than $\sim$10$^{13.4}$ M$_{\odot}$, 
%i.e. higher than radio-quiet quasars.

By contrast, Wake et al. (2008) 
%estimated the large-scale bias and halo mass for a sample of more
%luminous 2SLAQ radio AGNs.
%, i.e. we can assume that these
%objects are mainly FRII radio galaxies.
estimated an \textit{effective} DMH mass 
equal 10.3$\times$10$^{13}$ h$^{-1}$ M$_{\odot}$ for 2SLAQ quasars
at z = 0.55, by using the HOD.
This value is higher than typical halo masses derived for FRI objects
and \textit{Fermi}-Blazars. Indeed, this 
is quite expected 
%%Assuming that these
%%2SLAQ objects are mainly high power FRII radio galaxies
%%(P$_{1.4,GHz}\sim$ is 3 times higher 
%%than the previously cited samples), 
%%This difference in the halo mass 
given the different method (HOD) 
used to derive the DMH mass. 
%In fact, 
%while the \textit{effective} halo mass is, by definition, 
%weighted by the AGN HOD and the AGN number density, in our 
%method the bias is inferred by the
%clustering signal and weighted by the number of AGN pairs 
%(see Allevato et al. 2011).  
Instead, using the 
power-law best-fit parameters
quoted in Wake et al. (2008), we estimate a 
typical mass 
%for 2SLAQ radio AGNs significantly 
scaled to logM$_h$ = 13.6$\pm$0.13 h$^{-1}$ M$_{\odot}$.

In Fig. 3 we also show the typical DMH mass
of radio-loud SDSS quasars
at z = 0.28 (Krumpe et al. 2012) and
z = 1.3 (Shen et al. 2009). 
%These radio AGNs are not clearly classified as FRII radio galaxies
Given the high radio luminosities, we expect that these quasars 
are representative of FRII radio AGNs.
In detail, SDSS quasars at z=1.3 inhabit DMHs with
logM$_h$ = 13.4$\pm$0.20 h$^{-1}$ M$_{\odot}$, value which
is quite consistent with our result on \textit{Fermi}-FSRQs at similar redshift.

To summarize, we find that $\gamma$-ray Blazars 
and FRI-II radio AGNs reside in DMHs
of similar mass of the order of few times 10$^{13}$ h$^{-1}$ M$_{\odot}$
at low ($\sim 0.4$) and high ($\sim$1.2) redshift.
This results suggest that, in terms of clustering properties, Blazars and radio-loud galaxies 
are similar objects, supporting the unification scheme equating FRI-II radio AGNs as misaligned 
counterparts of Blazars (Urry \& Padovani 1995 and reference therein).

Similarly, measurements of the clustering of X-ray AGN
with moderate luminosity (L$_{bol}\sim10^{45-46}$erg s$^{-1}$)
show that they are located in dense environment, typical of 
galaxy groups ($10^{13-13.5}$h$^{-1}$M$_{\odot}$) at low 
($\sim$0.1) and high (1-2) redshift (e.g. Hickox et al. 2009, Cappelluti et al. 2010, Allevato et al. 2011, Krumpe et al. 2010, 2012, Mountrichas et al. 2012, Koutoulidis et al. 2013).
On the contrary, the hosting halo mass of $\gamma$-ray Blazars 
and FRI-II radio AGNs is an order of magnitude larger than
the typical mass of luminous quasars
(e.g. Croom et al. 2005, 2009, da Angela et al. 2008, Shen et al. 2009, Ross et al. 2009). In fact, 
several works have shown that 
luminous optically selected quasars are hosted by halos of roughly constant mass, a few times 10$^{12}$h$^{-1}$M$_{\odot}$, out to z$\sim$3.
%On the other hand, our results do not suggest a picture where
%FSRQs reside in very massive DMHs of the order of $\sim10^{14}$ h$^{-1}$ M$_{\odot}$
%as found for FRII radio AGNs. On the contrary, we inferred that
%BL Lacs and FSRQs inhabit the same environment.}
%with
%FRI galaxies and FSRQs with FRII galaxies, i.e. radio-loud AGN
%simply represent the misaligned counterparts of Blazars
%(Urry \& Padovani 1995 and reference therein).
%while FSRQs and FRIIs inhabit more massive halos with M$_h \sim 10^{14}$ 
%h$^{-1}$ M$_{\odot}$.

Studies of the cross correlation function between Blazars 
and large sample of galaxies will significantly reduce 
the uncertainties in the 2PCF allowing the modelling
of the signal with the halo occupation. The importance 
of using this method to derive more reliable estimate of the full halo mass 
distribution, is supported by 
Wake et al.'s results, which suggest 
more massive hosting halos for radio AGNs.

In terms of properties of Blazar hosting galaxies,
previous works have shown that the host
galaxies of Blazars are luminous giant ellipticals, 
regardless of intrinsic nuclear power.
Accordingly, we find that 2LAC Blazars are located in 
relatively large dark matter halos with DMH mass few $ 
\times 10^{13} h^{-1}M_{\odot}$, corresponding to the large galaxy groups or small clusters.
Additionally, Blazars also have same morphologies, luminosities and size as the host galaxies of FRI-II sources
(e.g., Urry et al. 2000; O'Dowd \& Urry 2005; Kotilainen et al. 2005).
Our results extend to the clustering properties 
(and then to the typical environment they live in)
the similarity between $\gamma$-ray Blazars and radio galaxies.

Following these results, 
we also expect that 
BL Lacs and FSRQs preferentially occupy the centre of DMHs
and host very massive BH ($>10^8$M$_{\odot}$),
as already pointed out for radio AGN 
(e.g. Wake et al. 2008, Mandelbaum et al. 2009, Smolcic et al. 2011,
Ghisellini et al. 2013).
The halo occupation will provide the full distribution 
of $\gamma$-ray AGNs among DMHs
and then the contribution from objects in central halos. 
%The importance of a more reliable estimate of the full mass 
%distribution of Blazar hosting halos by using the halo 
%occupation is supported by Wake et al.'s results, which suggest 
%more massive hosting halos for radio AGNs.
%compared with the direct 
%measurements of the Blazar auto-correlation. 
%and will allow us to estimate the mean fraction of $\gamma$-ray AGNs
%in central halos, which translates into the probability of having a contamination
%from Blazars to the expected signal of DM annihilation.
This is important in light of several projects that are currently 
looking for $\gamma$-ray emission from DM annihilation in 
the Galactic Centre, in dwarf galaxies and in galaxy clusters. 
%In fact, if DM is a thermal relic consisting of
%weakly interacting massive particles (WIMPs), then DM particle annihilation
%or decay can produce monochromatic $\gamma$-ray lines (photons)
%at $\sim$130 GeV.
%lines and 
%contribute to the diffuse $\gamma$-ray background
%(Ackermann et al. 2012).
%A $\gamma$-ray excess at this energy
%has been found in the \textit{Fermi}-LAT 
%(e.g. Bringmann et al. 2012, Tempel
%et al. 2012).
%Weniger 2012, Su \& Finkbeiner 2012). 
Galaxy clusters are the largest gravitationally bound 
structures in the Universe dominated by DM. Their large
DM content makes them interesting targets
for indirect detection of DM annihilation
%are the biggest nearby cosmological structures dominated by
%DM and are expected to be much better objects for that purpose
%because the DM annihilation signal from
%there should be amplified by a boost factor due to the existence
%of many DM subhalos 
(e.g. Nezri et al. 2012, Hektor et al. 2013).
In this light, the fact that $\gamma$-ray Blazars reside in clusters
and preferentially residing in central halos, 
%it is fundamental to understand the occupation of 
%Blazars in DMH in order to quantify
%the contamination from AGN to the putative detected signals
suggests a strong contamination from Blazars to the putative
detected signal of DM annihilation.
%Moreover, Fig. 2 shows that $\gamma$-ray selected Blazars are
%more biased and reside in more massive halos compared to bright quasars and moderate luminosity
%AGN (which are mainly X-ray selected). 
%But it is important to stress that we are not comparing $\gamma$-ray blazars with 
%optically and X-ray selected radio-quiet AGN, since radio-loud AGN are not excluded from
%the samples used in the works quoted in Fig. 2. 

%Studies of the cross correlation function between Blazars 
%and large sample of galaxies will significantly reduce 
%the uncertainties in the 2PCF \textbf{allowing the modelling
%of the signal with the halo occupation.}
%This approach
%will provide a more reliable estimate of the full mass 
%distribution of Blazar hosting halos. 
%The importance of a more reliable estimate of the full mass 
%distribution of Blazar hosting halos by using the halo 
%occupation is supported by Wake et al.'s results, which suggest 
%more massive hosting halos for radio AGNs.
%compared with the direct 
%measurements of the Blazar auto-correlation. 
%full dis- tribution of γ-ray AGNs among DMHs
%and will allow us to estimate the mean fraction of $\gamma$-ray AGNs
%in central halos, which translates into the probability of having a contamination
%from Blazars to the expected signal of DM annihilation.

\acknowledgments

VA, AF and wish to acknowledge Finnish Academy award, decision 266918.
This work has been supported by NASA grant NNX12AP22G.

\clearpage

\end{document}